# Rapid disappearance of a warm, dusty circumstellar disk

Carl Melis[1], B. Zuckerman[2], Joseph H. Rhee[3], Inseok Song[4], Simon J. Murphy[5], Michael S. Bessell[5]

[1]Center for Astrophysics and Space Sciences, University of California, San Diego, California 92093-0424, USA, [2]Department of Physics and Astronomy, University of California, Los Angeles, California 90095-1547, USA. [3]Department of Physics and Astronomy, California State Polytechnic University, Pomona, Pomona, CA 91768, USA, [4]Department of Physics and Astronomy, University of Georgia, Athens, GA 30602, USA, [5]Research School of Astronomy and Astrophysics, College of Mathematical and Physical Sciences, The Australian National University, Cotter Road, Weston Creek ACT 2611, Australia

**Stars form with gaseous and dusty circumstellar envelopes, which rapidly settle into disks that eventually give rise to planetary systems. Understanding the process by which these disks evolve is paramount in developing an accurate theory of planet formation that can account for the variety of planetary systems discovered so far. The formation of Earth-like planets through collisional accumulation of rocky objects within a disk has mainly been explored in theoretical and computational work in which post-collision ejecta evolution is typically ignored[1,2,3], although recent work has considered the fate of such material[4]. Here we report observations of a young, Sun-like star (TYC 8241 2652 1) where infrared flux from post-collisional ejecta has decreased drastically, by a factor of about 30, over a period of less than two years. The star seems to have gone from hosting substantial quantities of dusty ejecta, in a region analogous to where the rocky planets orbit in the Solar System, to retaining at most a meagre amount of cooler dust. Such a phase of rapid ejecta evolution has not been previously predicted or observed, and no currently available physical model satisfactorily explains the observations.**

TYC 8241 2652 1 (stellar parameters are reported in Table 1), was found as part of a survey to identify main sequence stars with excess emission at mid- and far-infrared wavelengths. To accomplish this goal we cross-correlated the Tycho-2 catalogue[5] with those of the *Infrared Astronomical Satellite* (*IRAS*), *AKARI*[6], and *Wide-field Infrared Survey Explorer* (*WISE*[7]) and performed our own observations using the Thermal-Region Camera Spectrograph[8] (T-ReCS) at the Gemini South telescope. Figure 1 and Table 2 show how the 11-μm excess emission of this source evolved from being a factor of ~30 times the stellar photosphere flux level before 2009 to being ~13 times the photospheric flux in mid-2009 and being barely detectable in 2010 (details regarding each measurement can be found in the notes to Table 2). The pre-2009 measurements indicate significant mid-infrared excess emission, and hence, that warm dusty material orbited in the star's inner planetary system (Figure 1 and Table 2). Remarkably, two epochs of *WISE* measurements show that the excess mid-infrared emission has all but disappeared leaving only a weak (~3 times the stellar photosphere) excess at a wavelength of 22 μm (Figure 1 and Table 2). We note that the two *WISE* epochs have a time separation of roughly six months and yet still report identical flux levels. Measurements made after the *WISE* epochs using the SpeX spectrograph at the NASA Infrared Telescope Facility[9,10,11], the Photodetector Array Camera and Spectrograph (PACS) for the *Herschel Space Observatory*[12] and again with T-ReCS are consistent with the *WISE* data (Figure 1; note especially the 2012 T-ReCS data), thus indicating that the mid-infrared emission from the dust orbiting this star has been consistently depleted to barely detectable levels since at least early 2010.

To determine the age of TYC 8241 2652 1 we obtained high-resolution optical spectra over four epochs from February 2008 to January 2009 with an echelle spectrograph mounted on the Siding Spring Observatory 2.3-m telescope. From these optical spectra we estimate the age of the system from the lithium content in the stellar photosphere, Galactic space motion, and rotational broadening of absorption lines;

details can be found in Supplementary Information. We adopt an age of ~10 Myr for TYC 8241 2652 1.

An important ingredient in understanding the vanishing mid-infrared excess emission toward TYC 8241 2652 1 is the initial state of its disk system. Given an age of ~10 Myr, the star could have been host to either an accreting protoplanetary disk rich in gas and dust or a second-generation debris disk formed from the collisions of rocky objects orbiting the star[13]. The absence of strong hydrogen Balmer Hα emission in our optical spectroscopic measurements indicates that the star was not undergoing accretion of hydrogen-rich material at any significant level[14] (see also Supplementary Information), and thus it is unlikely that such material was being transported inward to the star as would be expected in a system with an active protoplanetary accretion disk. Another argument against TYC 8241 2652 1 having a protoplanetary accretion disk in the two decades before 2009 lies with the *Herschel*/PACS measurements. The sensitive upper limits in the far-infrared robustly rule out the presence of a substantial reservoir of cold disk material typical of those seen in protoplanetary disks. We thus conclude that the dusty material orbiting TYC 8241 2652 1 is the result of the collisions of rocky objects.

To estimate the dust temperature and the fractional infrared luminosity ($L_{IR}/L_*$, where $L_*$ is the total stellar luminosity) of the dusty debris disk we fit optical and near-infrared measurements out to the $K_s$-band (2.1 μm) with a synthetic stellar atmosphere spectrum[15] along with a black body at 450 K (Figure 1) that models the pre-2009 epoch dust excess. Grains with a temperature of 450 K that are sufficiently large to radiate like black bodies at 10 μm and are situated in a disk optically thin to the stellar radiation field would orbit TYC 8241 2652 1 with a semi-major axis of ≈0.4 AU. From the blackbody fit we find that $L_{IR}/L_* ≈ 11\%$ (Figure 1); such a value is significantly greater than those found previously for stars with warm debris disks[16], but is less than that of the recently discovered ~60 Myr old V488 Per system[17]. A geometrically thin, flat dust disk (such as Saturn's rings or some circumstellar debris disks) cannot absorb 11% of the luminosity of TYC 8241 2652 1 (ref. 18) . To intercept such a large fraction of the incoming stellar light the disk must be geometrically thick or otherwise deformed into a non-flat shape. Such morphology could be suggestive of a substellar body that dynamically excites the dust particles, warps the disk, or both[19,20,21]. Fits to *WISE* and *Herschel*/PACS data allow a dust temperature only in the range $120 < T_{dust} < 250$ K, indicating cool grains that orbit TYC 8241 2652 1 at a distance of ~2 AU, and a fractional infrared luminosity of ~0.1%.

Given the luminosity of TYC 8241 2652 1 ($L_* \approx 0.7 L_\odot$, where $L_\odot$ is the solar luminosity), grains with radii of ~0.2 μm and smaller will be radiatively ejected from the disk system. Roughly $5 \times 10^{21}$ g in grains with radius of order 0.3 μm, which is slightly larger than the critical radius for radiative ejection, are required to produce the observed pre-2009 epoch infrared excess from TYC 8241 2652 1 (ref. 22). For a debris disk, the copious amounts of dust that were present suggest a system undergoing an active stage of terrestrial planet formation[16,23]. The excess emission detected by *WISE* requires roughly 4-5 times less mass in cool small (~0.3 μm) dust grains than that estimated for the grains at 450 K detected closer to the star in the pre-2009 epoch. To have no *WISE*-detectable signature of grains at 450 K requires that such dust contribute less than 0.1% to the total fractional infrared luminosity in the post-2009 epoch measurements.

It is desirable to develop a physical model that can explain the observed disappearance of the disk of dust grains at 450 K. In Supplementary Information we consider and reject models that rely on the disk material somehow being hidden from view thus resulting in the diminished flux. In lieu of these models, we explore others in which the disk material is physically removed from its pre-2009 epoch location. If the number of grains with radius $a$ that orbit the star follow a conventional $a^{-3.5}$ size distribution – and hence the fractional infrared luminosity scales like $a^{-0.5}$ (ref. 22) – we expect that removal of grains with radii up to ~1 mm would be required to eliminate the observational signature of dusty material orbiting at a separation of ~0.4 AU. For an $a^{-3.5}$ size distribution the diminished mid-infrared flux requires that the total mass of dust particles with $a$ less than ~1 mm located near 0.4 AU be smaller by a factor of ~100. A steeper grain size distribution with an exponent of –3.7 to –3.8 would require removal of grains with radii up to ~100 μm and would result in a reduction in the total grain mass by a factor of ~10.

Of the models explored in Supplementary Information, only the collisional avalanche[24] and runaway accretion[25] models are potentially viable, although each has its problems (details regarding these models and their shortcomings can be found in Supplementary Information). It is worth noting that both models benefit if the steeper grain size distribution is assumed and that modeling of other stars with warm debris disk indicates that such a steep power law slope may be present in such systems[26,27]. Clear identification of a physical model that can reproduce the observations will require modeling specific to the case of TYC 8241 2652 1 and its continued observation. Although the exact circumstances are not yet clear, this system has clearly undergone a drastic event that promises to provide unique insight into the process by which rocky planets form.

1. Wetherill, G. W. Formation of the Earth. *Annu. Rev. Earth Planet. Sci.* **18**, 205–256 (1990).

We thank Joel Kastner for advice regarding X-ray data and Michael Jura for suggesting the runaway accretion model. Based on observations obtained at the Gemini Observatory. This publication makes use of data products from the Two Micron All Sky Survey. This research made use of the SIMBAD and VizieR databases. C.M. acknowledges support from a LLNL Minigrant to UCLA and from the National Science Foundation. This work was supported in part by NASA grants to UCLA and University of Georgia.

The authors contributed equally to this work.

Reprints and permissions information is available at www.nature.com/reprints.

The authors declare no competing financial interests.

Correspondence and requests for materials should be addressed to C. M. (e-mail: cmelis@ucsd.edu).

**Table 1. Parameters of the star TYC 8241 2652 1**

| | |
|---|---|
| Right Ascension (R. A.) | 12 09 02.25 |
| Declination (Decl.) | -51 20 41.0 |
| Galactic Longitude (°) | 296.2104 |
| Galactic Latitude (°) | +10.9728 |
| Visual Magnitude | 11.5 |
| Spectral Type | K2 ± 1 |
| Effective Temperature (K) | 4950 ± 150 |
| Proper Motion in RA (mas yr$^{-1}$) | -34.1 ± 2.1 |
| Proper Motion in DEC (mas yr$^{-1}$) | -9.4 ± 2.0 |
| Heliocentric Radial Velocity (km s$^{-1}$) | 15 ± 1 |
| Lithium 6708 Å EW (mÅ) | 370 ± 10 |
| Hydrogen Balmer-$\alpha$ EW (Å) | 0.0 ± 0.1 |
| Ca II K emission core EW (Å) | 4.5 ± 0.5 |
| $v\sin i$ (km s$^{-1}$) | 10 ± 1 |
| Distance from Earth (pc) | 140 ± 20 (456 light years) |
| Galactic UVW Space Motions (km s$^{-1}$) | -12, -24, -7 |
| Age (Myr) | ~10 |

J2000 equinox Right Ascension and Declination are from the 2MASS catalogue. Galactic Longitude and Latitude are derived from the 2MASS R.A. and Decl. The spectral type and effective temperature is determined from line ratios[28] in the Siding Spring Observatory echelle spectra. Proper motion measurements are from the Tycho-2 catalog[5]. The radial velocity is measured from our Siding Spring echelle spectra by cross-correlating a target spectrum with a standard star spectrum of known radial velocity. Four epochs of radial velocity measurements

(UT 14 February 2008, 14 June 2008, 13 July 2008, and 12 January 2009) show no evidence for radial velocity variability within the measured errors (~1-2 km s$^{-1}$), ruling out any short orbital-period stellar companions to TYC 8241 2652 1. The radial velocity quoted in the table is the average of the four separate measurements. The listed hydrogen Balmer-$\alpha$, Ca II K core reversal emission, and Lithium 6708 Å equivalent widths (EW) are the average over the four Siding Spring echelle epochs and the uncertainty quoted is the standard deviation of those measurements. The velocity width ($v$ in $v$sin$i$, where $i$ is the angle of inclination of the stellar spin axis with respect to the line of sight towards Earth) was measured from the full-width at half-maximum depth (FWHM) of single absorption lines in the Siding Spring echelle spectra (which have intrinsic resolution element FWHM of ~13 km s$^{-1}$, a value we subtract in quadrature from the FWHM measured in the spectra). Velocities of the TYC 8241 2652 1 system (relative to the Sun) toward the center of our Milky Way galaxy, around the Galactic Center, and perpendicular to the Galactic plane (U, V, W) are calculated from Tycho-2 proper motions, our estimated photometric distance, and the optical echelle measured radial velocity – uncertainties on these values are roughly 2 km s$^{-1}$. See Supplementary Information for a discussion of the age.

**Table 2. Mid-infrared flux measurements of TYC 8241 2652 1**

| Observation Date(s) (Universal Time) | Instrument | Beam Size (arcseconds) | ≈10 μm Flux Density (mJy) | ≈20 μm Flux Density (mJy) |
|---|---|---|---|---|
| February - November, 1983 | IRAS | 45 x 270 (P.A. 132°) | 309 ± 31 | 224 ± 22 |
| May - November, 2006 | AKARI | 5.5 | 335 ± 14 | 315 ± 34 |
| May 6, 2008 | T-ReCS | 0.4 | 436 ± 44 | – |
| January 7, 2009 | T-ReCS | 0.4 | 164 ± 16 | – |
| January 8-10, 2010 | WISE | 6.1 | 12.8 ± 0.4 | 9.4 ± 0.8 |
| July 14-18, 2010 | WISE | 6.1 | 12.3 ± 0.5 | 9.2 ± 1.0 |
| May 1, 2012 | T-ReCS | 0.4 | 18 ± 6 | – |

Central wavelengths for each instrument are 12 and 25 μm for IRAS, 9 and 18 μm for AKARI, 10 μm for T-ReCS N-band imaging, and 11 and 22 μm for WISE. For each of the IRAS and WISE measurements, all available ancillary data products were examined to ensure reliability in the measured flux densities. T-ReCS observations were performed in clear, photometric conditions and are flux calibrated using consecutive observations of stars with known mid-infrared flux. For instruments other than IRAS, the quoted beam size is the point-spread function full-width at half-maximum. The large, irregular IRAS beam size is a result of the focal plane detector mask used, and the position angle (P.A.) is the orientation of this rectangular mask on the sky when IRAS observed TYC 8241 2652 1. A P.A. of 0° is North and a P.A. of 90° is East. For each of the IRAS, AKARI and WISE measurements, the satellite-measured stellar position agrees with the 2MASS position quoted in Table 1 to within the quoted errors (after taking into account the stellar proper motion listed in Table 1). For T-ReCS measurements, the stellar position is not absolutely determined in the observations but instead is determined relative to the calibration star that is observed immediately before or after observations of TYC 8241 2652 1. The position of the mid-infrared source detected toward TYC 8241 2652 1 on the T-ReCS detector array relative to the position of the calibration star on the detector array is a reflection of how close the detected source position is to the input position since the telescope slew precision is roughly 1 arcseond or better for small slews. In this manner we determine that, for each observation of TYC 8241 2652 1 with T-ReCS, the detected source lies within an arcsecond of the Table 1 stellar position. It is also noted that there is only one source detected in the 28.8" × 21.6" T-ReCS field-of-view and that each observation was sensitive enough to detect the photospheric flux level of TYC 8241 2652 1.

**Figure 1. Spectral Energy Distribution of TYC 8241 2652 1.** Measurements and the associated epoch (for mid- and far-infrared data) are indicated in the legend. The solid brown curve is a synthetic stellar photosphere[15] for a 4950 K effective temperature star that is fit to the optical and near-infrared data. The dotted line is a blackbody fit to the 12 and 25 $\mu$m IRAS excess data points - the temperature of this blackbody is 450 K and it suggests that roughly 11% of the optical and near-infrared starlight was being reprocessed into the mid-infrared by orbiting dust. The solid black line is the sum of the photosphere and the 450 K blackbody. Fitting a blackbody to the *WISE* and *Herschel* measurements suggests a dust temperature of roughly 200 K and a fractional infrared luminosity of 0.1%. Plotted flux density errors are one standard deviation. Some vertical error bars, for example those of the two earlier epochs of T-ReCS measurements, are smaller than the point sizes on the plot; for these measurements the uncertainty is comparable to or less than 10% of the corresponding measurement. Horizontal lines through each data point represent the filter full-width at half-maximum.

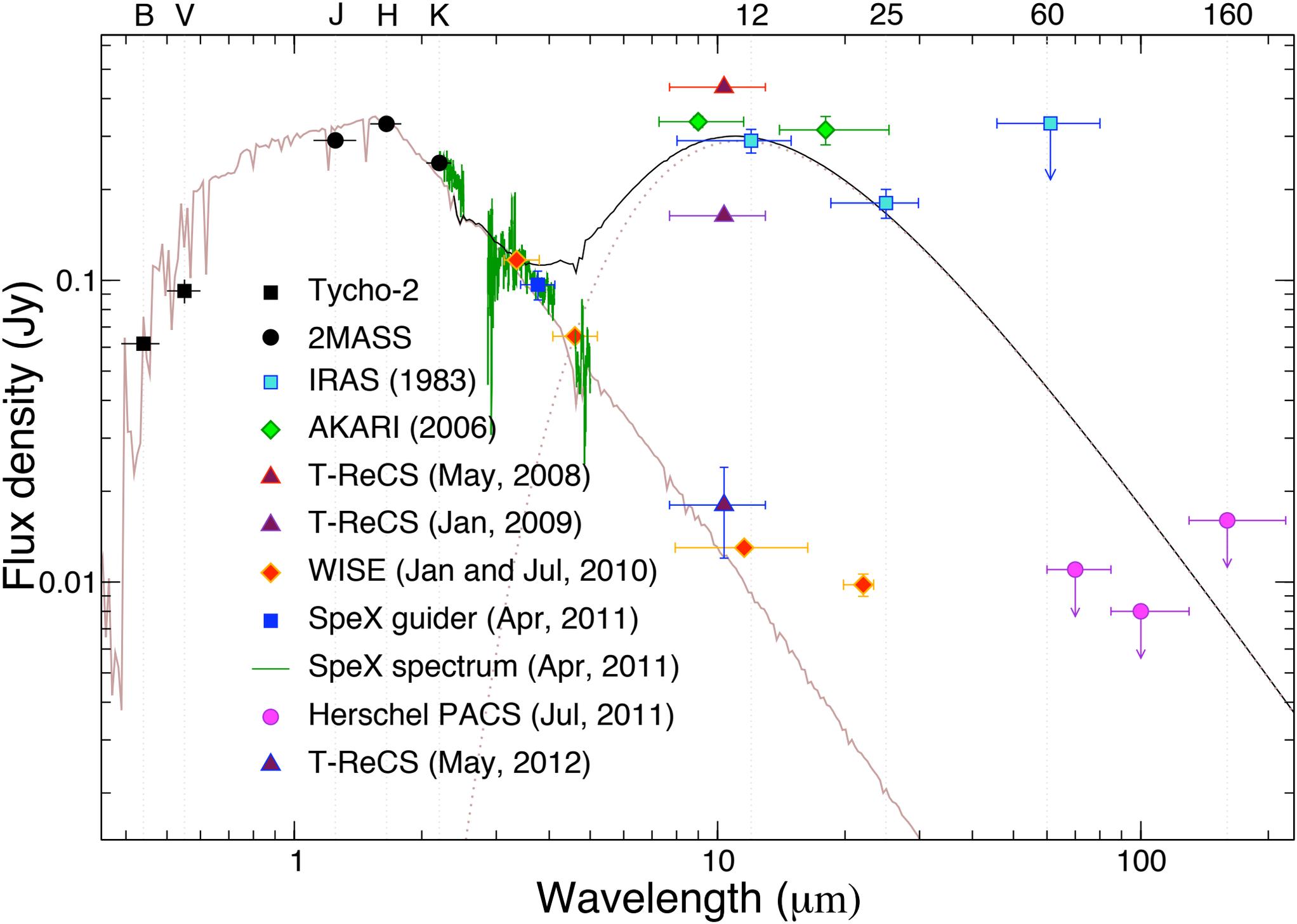

**Supplementary Information**

In this supplementary material we discuss in detail the age of TYC 8241 2652 1 and models for the disappearance of its dusty circumstellar disk.

**The Age of TYC 8241 2652 1**

Details regarding the method we use to determine an age for TYC 8241 2652 1 can be found in ref. 29. The lithium 6708 Å absorption feature is strongly detected in all spectra of TYC 8241 2652 1. Lithium content in a stellar atmosphere is mainly determined by the star's age and mass where, for stars of roughly solar mass, younger stars will show significant lithium content in their atmospheres and older stars will have depleted their lithium. From the measured lithium equivalent widths (EW, see Table 1) we are able to estimate an age of ~10 Myr[29]. Our calculated UVW Galactic space motions are consistent with the motions of stars in the solar neighborhood that are young (<100 Myr). The kinematics of TYC 8241 2652 1 are consistent with either of the Lower-Centaurus-Crux association[30,31] (10-20 Myr old[32,33]) or the TW Hydrae association (TWA, ~8 Myr old[32]). TYC 8241 2652 1 is located close in the plane of the sky to TWA 24 and shares similar space motions suggesting it could be a member of TWA. ROSAT All-Sky Survey and XMM-Newton X-ray observations of TYC 8241 2652 1 detect comparable flux levels near the stellar position but do not conclusively identify the star as the X-ray source. If the detected source is TYC 8241 2652 1, then the satellites suggest that the star has a fractional X-ray luminosity ($L_x/L_{bol} \sim 10^{-3}$) that is consistent with a young age for this system when compared to the X-ray flux from stars of known ages with similar effective temperatures[29]. Stellar activity indicative of a young star is also suggested by non-negligible stellar rotation[34], a filled Hα absorption line[29], and Ca II core reversal emission (all three parameters are reported in Table 1).

The age of ~10 Myr that is adopted for TYC 8241 2652 1 is consistent with the above considerations.

**Inner disk wall height variability**

Although there are reported cases of disk mid-infrared variability for very young stars with protoplanetary disks[35,36,37], none have mid-infrared variability similar to what we have observed. In these objects the inner rim of the protoplanetary disk changes in height casting a shadow on the outer disk material thus reducing the impinging stellar flux. This reduction in incident flux necessarily leads to a reduction in the re-radiated energy of the outer disk, which emits primarily in the mid- and far-infrared wavelengths due to its separation from its host star. The most variable source to date that has been modeled is ISO-52[36] where a factor of 2 change in its inner disk (mid-infrared) flux is observed. As with all other sources of this class, a reduction of flux in an outer disk must be accompanied by an increase of flux from the inner disk which shadows the outer disk. In the case of TYC 8241 2652 1 an "outer disk" is what would produce the pre-2009 epoch excess emission and an "inner disk" is an undetected hot dust component closer to the star. Taking the case of ISO-52 (which also has a "seesaw" point that is in the middle of the *Spitzer Space Telescope* InfraRed Spectrograph spectral range of 5-35 microns and hence has well-sampled emission variability at both long and short wavelengths[36]) we see that there is a 1.6:1 exchange in inner disk flux with outer disk flux[36]. That is to say, when the ISO-52 outer disk shows emission depressed by a factor of ≈1.25, the inner disk shows emission enhanced by a factor of ≈2. Applying this to TYC 8241 2652 1, we see that depression of "outer disk" emission by a factor of >30 suggests that any "inner disk" emission should be enhanced by a factor of >50; but the WISE 3.35 μm data taken simultaneously with the 11 and 22 μm data (which show the depressed flux) show no evidence for elevated flux. This makes a wall height change seem highly unlikely as being able to explain the observations.

The source [PZ99]J160421.7–213028 (hereafter [PZ99]J1604) has been observed to show a factor of 4 drop in mid-infrared flux[35], although little follow-up into this event has been performed. The drop in mid-infrared emission, if indeed real and not a data artifact, had by the time of the next mid-infrared measurements (5 months and 1 year later) recovered to its previous level[35]. Recent infrared measurements from *WISE* and by ref. 38 detect the source at the recovered mid-infrared emission level. [PZ99]J1604 is variable in the visible (0.55 nm), exhibiting up to factor of 6 drops in flux (in contrast with TYC 8241 2652 1; see Supplementary Figure 1). These visible light depressions are recurrent over long time baselines as seen in the ASAS project data[39]. [PZ99]J1604 is classified as a transition disk object[40] and is suspected to accrete at a very low level[41]. Its outer disk is especially massive ($\approx$10 $M_{Jup}$) for its age[41]. The highly variable visible dimming of the star in combination with the massive outer disk at [PZ99]J1604 argue for disk shadowing in a fortuitous observing geometry such that the star is dimmed as well. All of the above considerations clearly differentiate [PZ99]J1604 from TYC 8241 2652 1.

**Occultation of the disk by intervening line-of-sight material**

Models where the pre-2009 epoch warm disk material is occulted by optically opaque cooler dust located between Earth and the warm dust are not supported by the observations. There is no evidence for cool material in the *Herschel/PACS* observations. *WISE* 3.35 μm photometric measurements that detect the stellar photosphere flux level taken simultaneously with *WISE* measurements at 11 and 22 μm indicate that the underlying star is not occulted in the post-2009 epoch notwithstanding the factor of >30 optical extinction required in any occultation model that blocks out the mid-infrared emission. Similarly, optical monitoring of the source by the ASAS project[39] from 2000.9 to 2009.9 (a month before the first epoch *WISE* measurements were taken) indicate, outside of what appear to be a few spurious data points, that the star is stable (no significant or sustained optical dimming is seen; Supplementary Figure 1).

Another related model that can easily be ruled out is where the dust cloud is itself occulted by the host star. Such an event would require the disk to be in an edge-on orientation relative to our line-of-sight (a less than 20% chance alignment from purely geometric considerations). The perfect agreement of the two *WISE* epochs separated by six months time argues against such a model, but the most damning evidence against this model is that the solid angle in the sky subtended by the dust cloud must be >400 times larger than the solid angle of the star to reproduce the observed relative H-band and 11 μm flux densities in the pre-2009 epoch spectral energy distribution.

**X-ray Vaporization of the Dusty Disk**

Here we consider a scenario where a stellar X-ray flare emptied the dusty reservoir around TYC 8241 2652 1 by vaporizing its constituent dust grains. Young stars are known to undergo intense X-ray flare events, sometimes with luminosity comparable to or greater than the pre-flare bolometric luminosity of the host star[42,43]. A model for dust grain vaporization is presented in ref. 44. This model is developed to explore the interaction of Gamma-ray bursts with their surrounding medium, but the dust grain physics presented are applicable in a more general sense as they do not rely on any specific Gamma-ray burst properties. The only modification of this model for the case of TYC 8241 2652 1 is the shift of the relevant energy range for dust grain and photon interactions from the ultraviolet to X-rays (in the Gamma-ray burst scenario the X-rays are thought to be absorbed mostly by atomic and molecular hydrogen gas[44]); we explore the validity of this adaptation at the end of this subsection. We assume X-ray flare durations (where the luminosity over the time span is roughly constant) of 5-35 kiloseconds[43], refractory grain material composed of astrophysical silicates and graphite as the constituents of the disk system, that such grains absorb X-rays with energies typical of stellar X-ray flares[43] with efficiency of 100% (see below), and that these grains orbit TYC 8241 2652 1 with a semi-major axis of 0.4 AU. Inserting these parameters into Equations 11, 13, and 17 of ref. 44, we estimate that an X-ray flare with order of magnitude $10^2$-$10^3$ times the luminosity of the Sun is needed to vaporize enough grains to reduce the mid-infrared flux of the disk orbiting TYC 8241 2652 1 by a factor of ≈30. Since TYC 8241 2652 1 has luminosity of ≈0.7$L_\odot$, the required X-ray flare luminosity would be greater than $10^2$-$10^3$ times brighter than the star itself; to date, no such luminosity has been observed for stars of roughly Solar-mass, even for such young ages. A more critical problem is that X-rays are ineffective at heating grains much larger than 0.3 μm in size[45], whereas grains with radii up to ~1 mm or ~100 μm must be destroyed (see main text). This issue, combined with the high X-ray luminosity required by this model, makes it unlikely to explain the observed disk disappearance.

**Collisional Avalanche**

Compared with the above models, a potentially more plausible scenario invokes a runaway process known as a collisional avalanche[24,46]. The process begins with the release of numerous small dust grains from a collision involving a planetesimal-sized object in a pre-existing, gas-free dust disk. Small grains released by this collision are radiatively blown out of the disk system, impinge on other grains within the disk, and release from the impactee grains smaller than blowout sizes. For this process to reach a runaway state the disk system must be sufficiently dense such that a grain being driven from the system has a high likelihood of colliding with another (orbiting) grain in the disk and thus generating more outflowing grain fragments that can further generate more grain-damaging projectiles. Note that grains that are impacted need not be destroyed in a single collision – cratering impacts are sufficient to maintain the avalanche process – while grains up to 1 mm in size can be effectively chipped away to oblivion by numerous impacts that each remove a small amount of mass. As noted in the main text, if a grain size distribution steeper than $a^{-3.5}$ is present, then grains up to only ~100 μm must be destroyed thus requiring fewer cratering impacts per disk particle. Relative grain velocities of $>\sim 1$ km s$^{-1}$ are needed to whittle impactees into blow-out size grains[47]. In its pre-2009 dusty state, TYC 8241 2652 1 would certainly have satisfied the dustiness criterion as the initial model developed by ref. 24 was shown to operate for systems orders of magnitude less dusty.

The collisional avalanche model was originally developed for the case of outer planetary system debris disks and especially for the case of the disk orbiting β Pictoris[24]. Its specific predictions (e.g., the absolute timescale for the avalanche process to run) are intimately tied to the parameters of the β Pictoris system, but some extrapolation to the TYC 8241 2652 1 system can be made. A major difference between the two systems is the factor of ≈40 greater fractional infrared luminosity for TYC 8241 2652 1 in its pre-2009 state compared to β Pictoris. This level of dustiness likely reduces the total number of orbital timescales that it takes for the avalanche to propagate throughout the entire disk. The smaller semi-major axis of the disk orbiting TYC 8241 2652 1 could also enable a faster avalanche timescale. In this model, diminished infrared excess flux would not start to be evident until after the peak of avalanche activity (5 orbital periods for β Pictoris; Figures 2 and 4 in ref. 24) when the grain removal rate exceeds production of grains. One might then anticipate the TYC 8241 2652 1 disk flux would begin to diminish after a delay of a few orbital periods from the hypothesized initial planetesimal break-up. To agree with the observed removal timescale of the TYC 8241 2652 1 disk, grains produced by the avalanche process need to be radiatively driven sufficiently far from their originating position such that they do not contribute to the disk excess infrared emission in ~2.5 years. For a disk orbiting TYC 8241 2652 1 with semi-major axis of ≈0.4 AU, 2.5 years corresponds to roughly 10 orbital periods.

It is desirable to explore a collisional avalanche for the specific case of TYC 8241 2652 1 given its potential to explain the observations. Until such focused modeling is performed, it is difficult to make robust predictions about the final outcome of the avalanche process or the likelihood of its occurrence. Compared to the nominal case given by β Pictoris, the probability of observing this process in a disk as dense as that orbiting TYC 8241 2652 1 in its pre-2009 epoch state is quite high[24]. This probability depends on, in addition to the dustiness of the disk system, how much material is released by the initial planetesimal collision. Exploration of the dependency of the collisional avalanche process as a function of initial released dust mass and size distribution will allow a more robust characterization of the likelihood of having a collisional avalanche occur in TYC 8241 2652 1's dusty disk and its capability to explain the available (or future) observations. Such a process is likely cyclical, in that the disk seen around TYC 8241 2652 1 in the pre-2009 epoch will probably regenerate if it indeed was emptied by a collisional avalanche. The regeneration timescale depends on the largest size grains in the disk that are depleted by the avalanche process – if grains up to 1 mm are depleted in the disk, then regeneration could occur after roughly two decades.

**Runaway Accretion**

Another mechanism that might reproduce the observed disk disappearance timescale is runaway accretion of the dusty disk caused by aerodynamic drag induced on the disk grains by gas within the disk[25,48]. The physical mechanisms describing this process are outlined in ref. 25 and the results indicate that material in an orbiting disk can be accreted at rates that are hundreds of times faster than can be achieved by Poynting-Robertson drag (PR) alone. It is worth noting that the PR timescale being considered for the white dwarf disks in refs. 25 and 48 is orders of magnitude faster than that for the grains that must be removed from the disk orbiting TYC 8241 2652 1.

Taking the TYC 8241 2652 1 PR timescale into account, and assuming that the physics is similar between the white dwarf and TYC 8241 2652 1 disks[25,48], runaway accretion rates are sufficient to explain the disappearance of grains with size up to ~100 μm or perhaps even ~1 mm on timescales of a few years.

While many youthful main sequence stars are known to host dusty debris disks, only a handful of stars with ages comparable to TYC 8241 2652 1 are also known to host gaseous disks[49,50,51,52]. The gas in these disks is detected by radio observations that are sensitive only to gas at large (~100 AU) semi-major axes. It is not known whether inner planetary system debris disks also host gaseous material. For the TYC 8241 2652 1 pre-2009 epoch disk to be subject to the runaway accretion mechanism, it is necessary for it to have at least as much gas as dust in the system (gas-to-dust mass ratio of unity) and probably as much as ten times more gas than dust for the mechanism to operate effectively[25,48]. If we assume a constant accretion rate over the duration of the disk disappearance (taken here to be roughly 2 years), a gas-to-dust ratio of 10, and that there are roughly $10^{23}$ grams of 1 mm sized grains corresponding to the case of a conventional grain size distribution[22], then the accretion rate would be about $10^{-9}$ to $10^{-8}$ $M_\odot$ yr$^{-1}$. The origin of the gas necessary to drive the runaway accretion process is a major uncertainty of this model. One possibility is that there is residual gaseous material left over from the protoplanetary disk. Such gaseous material would be dominantly hydrogen in composition which leads to a testable prediction. The accretion rate calculated above for hydrogen-dominated material would result in Hydrogen Balmer-α emission from the accretion shock[53]. No Balmer-α emission is detected in any of the four epochs of our spectroscopy (see Table 1 in the main text), indicating that active accretion of a hydrogen-rich gas and dust mixture is unlikely to be occurring. Another possibility is that the gaseous material is not remaining from the star-formation event, but that it results instead from the giant impact event that generated the dusty material seen to orbit TYC 8241 2652 1 in the pre-2009 epoch. As discussed in refs. 26, 54, and 55, it is possible to generate a significant amount of gaseous rocky vapor by giant impacts. Such rocky vapor in favorable environments (like that supplied by the pre-2009 epoch TYC 8241 2652 1 disk) could survive for timescales comparable to those of extremely dusty, warm disk systems[16,26]. Thus, perhaps rocky vapor sufficient to drive the runaway accretion process still survives from the giant impact event that generated the pre-2009 epoch TYC 8241 2652 1 dusty disk.

A test of the runaway accretion model can be made through the continuum excess luminosity released by the accreted particles[56]. Following the veiling measurement methodology of ref. 57, we determine that there is no detectable veiling in the echelle spectra of TYC 8241 2652 1, even in the January 2009 epoch taken close in time to the 2009 T-ReCS measurement that shows the mid-infrared emission diminished by a factor of roughly 4 (which implies that accretion would have already begun). We arrive at a veiling upper limit of r ($=F_{accretion}/F_{photosphere}$) < 0.1. With this value, and following the methodology of ref. 11, we determine that for TYC 8241 2652 1 the accretion rate is limited to < ~$10^{-11}$ $M_{\odot}$ yr$^{-1}$. Taking into account model-dependent uncertainties (which can be up to a factor of 10 in magnitude), we relax this upper limit to < ~$10^{-10}$ $M_{\odot}$ yr$^{-1}$. Regardless, it appears as though the veiling constraint does not leave much room for accretion at the rate derived in the preceding paragraph if TYC 8241 2652 1 accretes disk material with a conventional grain size distribution ($a^{-3.5}$) at a constant rate. If one assumes a steeper grain size distribution (e.g., $a^{-3.8}$), then less mass need be removed from the disk over the 2 year timescale. The resulting constant accretion rate of ~$10^{-10}$ $M_{\odot}$ yr$^{-1}$ is roughly consistent with the observationally determined accretion limit from veiling. This leaves open the possibility that a runaway accretion process did act to deplete the disk orbiting TYC 8241 2652 1. The runaway accretion process probably would not be cyclical like the collisional avalanche, and it would likely take a greater amount of time to replenish the dusty reservoir orbiting the star (note that this depends on the models as laid out in their original papers – these estimates might change for the specific case of TYC 8241 2652 1). If runaway accretion is the dominant mechanism that acted to remove the dusty material orbiting TYC 8241 2652 1, then this would imply that significant gaseous material exists and could interact with post-collision ejecta and any forming protoplanets.

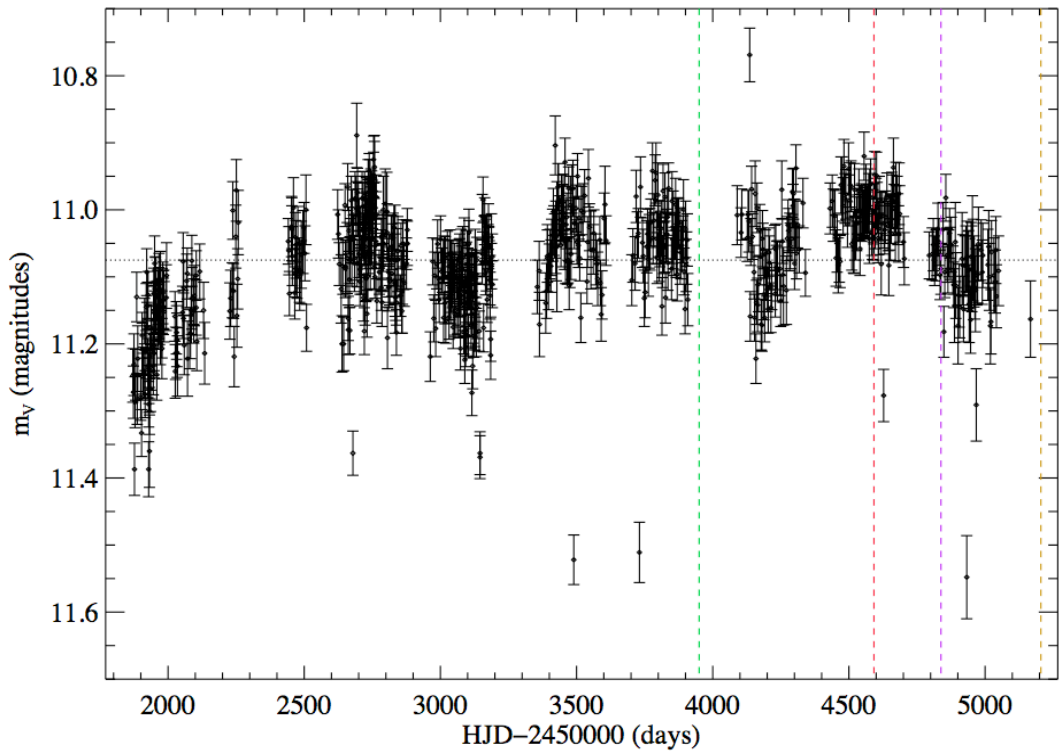

**Supplementary Figure 1. ASAS *V*-band measurements of TYC 8241 2652 1.** Data points and associated uncertainties were extracted from the ASAS project[39]. The abscissa is the heliocentric Julian date (spanning from roughly 2000.9 to 2009.9) while the ordinate is apparent visual (0.55 nm) magnitude. Catalog entries with valid measurements (value other than 29.999 or 99.999) for all aperture sizes were considered in constructing the light curve. The aperture that yielded the lowest measurement uncertainty is plotted for each epoch. The horizontal dotted line is the median of all plotted values. The colored vertical dashed lines correspond to the various epochs of mid-infrared measurements (from left to right, respectively): *AKARI* (green), first and second T-ReCS (red and purple), and first *WISE* (gold).